\documentclass [a4paper,12pt]{article}
\usepackage [T1]{fontenc}
\usepackage [cp1250]{inputenc}
\usepackage {amsmath}
\usepackage {amssymb}
\usepackage{pdflscape}
\usepackage {amsfonts}
\usepackage {amsthm}
\usepackage {longtable}
\usepackage {float}
\usepackage {cite}
\usepackage{bm}
\usepackage{multirow}
\usepackage{booktabs}

\newcommand{\gt}{>}

\begin{document}

\begin{center}\textbf{ROBUST VALIDATION OF APPROXIMATE 1-MATRIX FUNCTIONALS WITH FEW-ELECTRON HARMONIUM ATOMS}
\end{center}
\vspace{20pt}

\noindent Jerzy Cioslowski* $^{a)}$, Mario Piris$^{b,c)}$, and Eduard Matito$^{b,c)}$

\vspace{.04\textheight}

\noindent $^{a)}$ Institute of Physics, University of Szczecin, Wielkopolska 15, \\ 70-451 Szczecin, Poland
\vspace{16pt}

\noindent $^{b)}$ Kimika Fakultatea, Euskal Herriko Unibertsitatea (UPV/EHU), and \\ Donostia International Physics Center (DIPC),
20018 Donostia, \\ Euskadi, Spain
\vspace{16pt}

\noindent $^{c)}$ IKERBASQUE, Basque Foundation for Science, 48011 Bilbao, Euskadi, Spain

\vspace{25pt}
{\small{\noindent\underline{Abstract:} A simple comparison between the exact and approximate correlation components $U$ of the electron-electron repulsion energy of several states of few-electron harmonium atoms with varying confinement strengths provides a stringent validation tool for 1-matrix functionals. The robustness of this tool is clearly demonstrated in a survey of 14 known functionals, which reveals their substandard performance within different electron correlation regimes.  Unlike spot-testing that employs dissociation curves of diatomic molecules or more extensive benchmarking against experimental atomization energies of molecules comprising some standard set, the present approach not only uncovers the flaws and patent failures of the functionals but, even more importantly, also allows for pinpointing their root causes.  Since the approximate values of $U$ are computed at exact 1-densities, the testing requires minimal programming and thus is particularly suitable for rapid screening of new functionals.}

\vspace{50pt} \noindent {\small{$\overline{\textrm{*To} }\:
\textrm{whom all the correspondence should be addressed}$}}

\newpage
\normalsize\noindent\underline{I. INTRODUCTION}\vspace{8pt}

The present implementations of the density matrix functional theory (DMFT), in which the one-electron reduced density matrix (a.k.a.~1-matrix) $\Gamma(1',1)$ plays the central role \cite{a,b,c,d,e,f}, are not on par in accuracy with the wavefunction-based methods of quantum chemistry and the parameterizations of the relevant functionals are far less sophisticated than those of their density functional theory (DFT) counterparts.  Paradoxically, this state of affairs, coupled with the wealth of constraints that have to be satisfied by the 1-matrix functionals \cite{g,h,i,j}, bodes well for mathematical rigor and sound physical basis of approximate formulations of DMFT that are to emerge in the near future.  Another favorable circumstance stems from the fact that, unlike in the case of DFT, the homogeneous electron gas does not provide a convenient starting point for construction and testing of DMFT-based approaches to the electron correlation problem. Consequently, more realistic model systems have to be employed in its place.

The two-electron harmonium atom, described by the nonrelativistic Hamiltonian \cite{k,l}
\begin{eqnarray}\label{1}
\hat H =\frac{1}{2} \, \sum_{i=1}^N(-\hat\nabla_i^2+\omega^2\,r_i^2)
+\sum_{i>j=1}^N |\vec r_i -\vec r_j|^{-1}  \quad 
\end{eqnarray}
 with $N=2$, is an archetype of quasi-solvable systems of relevance to electronic structure theory.  As such, it has been repeatedly used in calibration and benchmarking of approximate electron correlation methods, especially those involving DFT \cite{m,n,o,p,q,r,s}.  However, due to the simplicity of the expression for the ground-state energy of a  two-electron system in terms of its 1-matrix, the two-electron harmonium atom is of no interest to the developers of DMFT.  In contrast, there are multiple reasons for which its congeners with $ N \gt 2$ are ideal model systems in this context \cite{s1}.  First of all, such atoms offer unlimited continuous tunability of extents and relative strengths of the dynamical and nondynamical electron correlation effects.  Thus, for large values of $\omega$, they are weakly correlated systems that, depending on $N$ and the electronic state, are described by either one or a few Slater determinants \cite{t}.  Conversely, at the $\omega \to 0$ limit of a vanishing confinement strength, their electrons  exhibit complete spatial  localization and exclusively nondynamical correlation \cite{k,l,u,v,w,x}.  The absence of a sudden Wigner crystallization allows for smooth interpolation between these weak- and strong-correlation regimes \cite{x1}.  Second, lacking the electron-nucleus cusps in their wavefunctions, all harmonium atoms are amenable to calculations involving basis sets of explicitly correlated Gaussian functions that yield highly accurate electronic properties \cite{y}.  Indeed, energies of several electronic states of the three- and four-electron harmonium atoms are presently known within ca.~1 [$\mu$hartree] for arbitrary values of $\omega$ \cite{y1,y2}.  The respective 1-matrices and individual energy components are also available from such calculations.  Third, exact asymptotics of these electronic properties are available at both the weak- and strong-correlation limits \cite{t,u,v,w,x,y3,z}, facilitating imposition of well-defined constraints upon the DMFT expressions for the total energy.

In summary, many-electron harmonium atoms are well suited for robust validation of approximate 1-matrix functionals.  Although such a validation can be carried out in multiple ways, the most convenient (and computationally least expensive) approach relies on comparison between the exact and approximate correlation components of the electron-electron repulsion energy $W$. In Coulombic systems, partitioning of $W$ parallels that of the 2-matrix \cite{aa,bb}.  Thus, $W$ comprises the direct (Coulomb) and exchange components, given by the expressions
\begin{eqnarray}\label{2}
J = \frac{1}{2} \, \int \hspace{-6pt} \int \Gamma(1,1) \, \Gamma(2,2) \, |\vec r_1-\vec r_2|^{-1} 
\; d1 \, d2
\end{eqnarray}
and
\begin{eqnarray}\label{3}
K = -\, \frac{1}{2} \, \int \hspace{-6pt} \int \Gamma(1,2) \, \Gamma(2,1) \, |\vec r_1-\vec r_2|^{-1} 
\; d1 \, d2 \quad,
\end{eqnarray}
respectively, and the remainder $U$ due to electron correlation.  This reminder, which originates from the diagonal part of the 2-cumulant matrix, is the only contribution to the total energy not given by an explicit functional within DMFT.  Consequently, evaluation of the difference between the (almost) exact $U$, obtained either from highly accurate calculations or asymptotic expressions, and its approximate counterpart computed with the exact $\Gamma(1',1)$  (originating from the same source as the exact $U$) offers a rigorous test of the performance of a given functional that involves minimal (if any) programming as no optimization of the 1-matrix is required.
 
In this paper, we employ the aforedescribed validation tool in a survey of the majority of the currently known 1-matrix functionals.  The origins of their flaws and patent failures are elucidated.

\vspace{16pt}

\normalsize\noindent\underline{II. THE INVENTORY OF APPROXIMATE  1-MATRIX FUNCTIONALS}\vspace{8pt}

The 1-matrix functionals tested in this study fall into two broad categories.  The first of them encompasses expressions for $U$ that involve only the exchange integrals 
$\{ K_{pq} \}$, where $K_{pq} =\langle \psi_p(1) \psi_q(2)|r_{12}^{-1}|\psi_q(1) \psi_p(2) \rangle $, computed with the natural spinorbitals $\{ \psi_p \}$ (NOs) and the respective occupation numbers $\{n_p\}$  \cite{cc}.  This category includes:
\vspace{6pt}

\noindent 1. The functional introduced by M\"uller \cite{f1} and later elaborated by Buijse and Baerends \cite{f2},
\begin{eqnarray}\label{4}
U_{MBB}= \frac{1}{2} \, \sum_{pq} \, (n_p \, n_q - \sqrt{n_p \, n_q}\,) \, K_{pq}  \quad  ;
\end{eqnarray}

\noindent 2. The functional  of Goedecker and Umrigar \cite{f3,f4},
\begin{eqnarray}\label{5}
U_{GU}=\frac{1}{2} \, \sum_{p \neq q} \, (n_p \, n_q - \sqrt{n_p \, n_q}\,) \, K_{pq}  \quad  ;
\end{eqnarray}

\noindent 3. The functional  of Cs{\'a}nyi and Arias \cite{f5},
\begin{eqnarray}\label{6}
U_{CA}= -\, \frac{1}{2} \, \sum_{pq} \, \sqrt{n_p \, n_q \, (1-n_p) \, (1-n_q)} \; K_{pq}  \quad  ;
\end{eqnarray}

\noindent 4. The functional  of Cs{\'a}nyi, Goedecker, and Arias \cite{f6},
\begin{eqnarray}\label{7}
U_{CGA}= \frac{1}{4} \, \sum_{pq} \, \left[ n_p \, n_q -  \sqrt{n_p \, n_q \, (2-n_p) \, (2-n_q)}
\, \right] K_{pq}  \quad  ;
\end{eqnarray}

\noindent 5. The BBC1 functional \cite{f7,f8},
\begin{eqnarray}\label{8}
U_{BBC1}= U_{MBB} +  \sum_{p \neq q} \, \theta(1-2 n_p) \, \theta(1-2 n_q) \, \sqrt{n_p \, n_q} \;
K_{pq}  \quad  ,
\end{eqnarray}
where $\theta(t)$ is the Heaviside step function;

\noindent 6. The BBC2 functional \cite{f7,f8},
\begin{eqnarray}\label{9}
U_{BBC2}= U_{BBC1} + \, \frac{1}{2} \, \sum_{p \neq q} \, \theta(2 n_p-1) \, \theta(2 n_q-1) \,
(\sqrt{n_p \, n_q}- n_p \, n_q ) \, K_{pq}  \quad  ;
\end{eqnarray}

\noindent 7. The functional of Marques and Lathiotakis \cite{f9},
\begin{eqnarray}\label{10}
U_{ML}= - \, \frac{1}{2} \, \sum_{pq} \, n_p \, n_q 
\left( \frac{a_0+a_1 \, n_p \, n_q}{1+b_1 \, n_p \, n_q} -1 \right)  K_{pq}  \quad  ,
\end{eqnarray}
where $a_0=126.3101$, $a_1=2213.33$, and $b_1=2338.64$;

\noindent 8. The functional of Marques and Lathiotakis corrected for self-interaction \cite{f9,f10},
\begin{eqnarray}\label{11}
U_{ML-SIC}= - \, \frac{1}{2} \, \sum_{p \ne q} \, n_p \, n_q 
\left( \frac{a_0+a_1 \, n_p \, n_q}{1+b_1 \, n_p \, n_q} -1 \right)  K_{pq}  \quad  ,
\end{eqnarray}
where $a_0=1298.780$, $a_1=35114.4$, and $b_1=36412.2$;

\noindent 9. The power functional \cite{f11,f12,f13},
\begin{eqnarray}\label{12}
U_{\lambda} = \frac{1}{2} \, \sum_{pq} \, \left[ n_p \, n_q - (n_p \, n_q)^{\lambda} \, \right]  K_{pq}  \quad  .
\end{eqnarray}
\vspace{6pt}

\noindent Although the sums in Eqs.~(\ref{4})-(\ref{12}) run over (almost) all pairs of NOs, only those with parallel spins contribute.  Despite their obvious shortcoming of neglecting correlation of electrons with antiparallel spins, these expressions have been employed in calculations on both closed- and open-shell systems \cite{f13,f14,f15,f16,f17}.

The genuine "JKL-only" \cite{cc,g0} functionals of the PNOF (Piris natural orbital functional) family obtain from model reconstructions of the 2-cumulant matrix and thus include the Coulomb integrals 
$\{ J_{pq} \}$, where $J_{pq} =\langle \psi_p(1) \psi_q(2)|r_{12}^{-1}|\psi_p(1) \psi_q(2) \rangle $, in the approximate formulae for $U$, alleviating the aforementioned deficiency.  Unfortunately, the assumption of NOs possessing pairwise-identical spatial parts limits the applicability of the PNOF functionals to spin-unpolarized systems and species with all-parallel spins, each of those two cases requiring a distinct expression for $U$.  For the sake of reader's convenience, these expressions are compiled below: 
\vspace{6pt}

\noindent 1. The PNOF1 functional \cite{g1,g2}:
\begin{eqnarray}\label{13}
U_{PNOF1}= \frac{1}{2} \, \sum_{p \ne q} \, \left[ n_p \, n_q - \nu_p \, \nu_q \, \sqrt{h_p \, h_q}
+(2 \, \eta_p \, \eta_q-1) \, \sqrt{n_p \, n_q} \, \right] K_{pq} \quad  
\end{eqnarray}
for the spin-unpolarized systems and 
\begin{eqnarray}\label{14}
U_{PNOF1}=\frac{3}{8} \, \sum_p \, (n_p^2-n_p) \,  K_{pp} \quad  
\end{eqnarray}
for the high-spin ones;

\noindent 2. The PNOF2 functional \cite{g3}:
\begin{eqnarray}\label{15}
&& \hspace{-10pt} U_{PNOF2}= \frac{1}{2} \, \sum_p \, (n_p-n_p^2) \, K_{pp} \nonumber \\
&& \hspace{-10pt} - \, \frac{1}{2} \, \sum_{p \ne q} \, \big[ \nu_p \, \nu_q \, h_p \, h_q 
+ (s_F^{-1}-1) \, (\nu_p \, \eta_q \, h_p \, n_q + \eta_p \, \nu_q \, n_p \, h_q) 
+ \eta_p \, \eta_q \, n_p \, n_q \, \big] J_{pq} \nonumber \\
&&  \hspace{-10pt} + \, \frac{1}{2} \, \sum_{p \ne q} \,  \Big[ n_p \, n_q 
- \big( \nu_p \, \sqrt{h_p}-\eta_p \, \sqrt{n_p}\, \big) \, 
\big( \nu_q \, \sqrt{h_q}-\eta_q \, \sqrt{n_q}\, \big) \nonumber \\
&& \hspace{200pt} +(2 \,\eta_p \, \eta_q-1) \, \sqrt{n_p \, n_q} \, \Big] K_{pq} \quad 
\end{eqnarray}
for the spin-unpolarized systems and 
\begin{eqnarray}\label{16}
U_{PNOF2}= -\, \frac{1}{2} \, \sum_{pq} \, \big[ \nu_p \, \nu_q \, h_p \, h_q 
+ (s_F^{-1}-1) \, (\nu_p \, \eta_q \, h_p \, n_q + \eta_p \, \nu_q \, n_p \, h_q) \nonumber \\
+ \eta_p \, \eta_q \, n_p \, n_q \, \big] (J_{pq}-K_{pq}) \quad  
\end{eqnarray}
for the high-spin ones;

\noindent 3. The PNOF3 functional \cite{g4}:
\begin{eqnarray}\label{17}
&& \hspace{-30pt} U_{PNOF3}=\frac{1}{2} \, \sum_p \, (n_p-n_p^2) \, K_{pp} \nonumber \\
&&  \hspace{-30pt} - \, \frac{1}{4} \, \sum_{p \ne q} \, \big[ \nu_p \, \nu_q \, h_p \, h_q 
+ (s_F^{-1}-1) \, (\nu_p \, \eta_q \, h_p \, n_q + \eta_p \, \nu_q \, n_p \, h_q) 
+ \eta_p \, \eta_q \, n_p \, n_q \, \big] J_{pq} \nonumber \\
&& \hspace{-30pt} + \, \frac{1}{2} \, \sum_{p \ne q} \,  \Big[ n_p \, n_q 
-\nu_p \, \eta_q \, \sqrt{h_p \,  n_q} -\eta_p \, \nu_q \, \sqrt{n_p \,  h_q} 
+(2 \,\eta_p \, \eta_q-1) \, \sqrt{n_p \, n_q} \, \Big] K_{pq} \nonumber \\ \quad 
\end{eqnarray}
for the spin-unpolarized systems and 
\begin{eqnarray}\label{18}
U_{PNOF3}= 0 \quad  
\end{eqnarray}
for the high-spin ones;

\noindent 4. The PNOF4 functional \cite{g5}:
\begin{eqnarray}\label{19}
&&\hspace{-10pt} U_{PNOF4}=  \frac{1}{2} \, \sum_p \, (n_p-n_p^2) \, K_{pp} \nonumber \\
&&\hspace{-10pt} - \, \frac{1}{2} \, \sum_{p \ne q} \, \big[ \nu_p \, \nu_q \, h_p \, h_q 
+ (s_F^{-1}-1) \, (\nu_p \, \eta_q \, h_p \, n_q + \eta_p \, \nu_q \, n_p \, h_q) 
+ \eta_p \, \eta_q \, n_p \, n_q \, \big] J_{pq} \nonumber \\
&&\hspace{-10pt} + \, \frac{1}{2} \, \sum_{p \ne q} \,  \Bigg( \nu_p \, \nu_q \, 
\big( h_p \, h_q-\sqrt{h_p \, h_q} \, \big) 
+\eta_p \, \eta_q \, \big( n_p \,  n_q + \sqrt{n_p \, n_q} \, \big) \nonumber \\
&&\hspace{40pt} + \, s_F^{-1} \, \Big[ 1-s_F-\sqrt{1+ \frac{s_F}{h_p \, n_q} \,  (n_p-n_q)} \; \Big]
\nu_p \, \eta_q \, h_p \, n_q  \nonumber \\
&&\hspace{40pt} + \, s_F^{-1} \, \Big[ 1-s_F-\sqrt{1+ \frac{s_F}{n_p \, h_q} \,  (n_q-n_p)} \; \Big]
\eta_p \, \nu_q \, n_p \, h_q  \Bigg) K_{pq} \quad  
\end{eqnarray}
for the spin-unpolarized systems and 
\begin{eqnarray}\label{20}
U_{PNOF4}= U_{PNOF2} \quad  
\end{eqnarray}
for the high-spin ones;

\noindent 5. The PNOF6 functional \cite{g6}:
\begin{eqnarray}\label{21}
&&\hspace{-48pt} U_{PNOF6}=   \frac{1}{2} \, \sum_p \, (n_p-n_p^2) \, K_{pp} \nonumber \\
&&\hspace{-5pt} - \, \frac{1}{2} \, \sum_{p \ne q} \, \big[  \exp{(-2s_F)} \, 
\nu_p \, \nu_q \, h_p \, h_q + B_{pq} \, \nu_p \, \eta_q \nonumber \\
&&\hspace{25pt} + B_{qp} \, \eta_p \, \nu_q
+ \exp{(-2s_F)} \, \eta_p \, \eta_q  \, n_p \, n_q \, \big] J_{pq} \nonumber \\
&&\hspace{-5pt} + \, \frac{1}{2} \, \sum_{p \ne q} \,  \Bigg( \nu_p \, \nu_q \,  
\big[ \exp{(-2s_F)} \, h_p \, h_q -\exp{(-s_F)} \, \sqrt{h_p \, h_q} \; \big] \nonumber \\
&&\hspace{40pt} + \, \eta_p \, \eta_q \, 
\big[ \exp{(-2s_F)}\, n_p \, n_q+\exp{(-s_F)} \, \sqrt{n_p \, n_q} \; \big] \nonumber \\
&&\hspace{40pt}  + \, \nu_p \, \eta_q \, 
\Big[ B_{pq}-\sqrt{(n_p \, h_q+B_{pq}) \, (h_p \, n_q+B_{pq})} \; \Big]  \nonumber \\
&&\hspace{40pt} + \, \eta_p \, \nu_q \,
\Big[ B_{qp}-\sqrt{(n_p \, h_q+B_{qp}) \, (h_p \, n_q+B_{qp})} \; \Big]  \Bigg) K_{pq} \quad  
\end{eqnarray}
for the spin-unpolarized systems and 
\begin{eqnarray}\label{22}
&&\hspace{-40pt} U_{PNOF6}= - \, \frac{1}{2} \, \sum_{pq} \, \big[  \exp{(-2s_F)} \, 
\nu_p \, \nu_q \, h_p \, h_q + B_{pq} \, \nu_p \, \eta_q \nonumber \\
&&\hspace{65pt} + B_{qp} \, \eta_p \, \nu_q
+ \exp{(-2s_F)} \, \eta_p \, \eta_q  \, n_p \, n_q \, \big]  (J_{pq}-K_{pq})  \quad  
\end{eqnarray}
for the high-spin ones, where
\begin{eqnarray}\label{23}
&\hspace{-65pt} B_{pq}=\Big(- s_F^2 \, \exp{(-2s_F)} + \sum\limits_p \eta_p \, \big[n_p \,  h_p
+\exp{(-2s_F)} \, n_p^2 \, \big] \, \Big)^{-1} \nonumber \\ 
&\hspace{-10pt} \times \, \big[ n_p+\exp{(-2s_F)} \, (h_p-s_F) \big]
\big[ h_q+\exp{\left(-2s_F\right)} \, (n_q-s_F) \big] \, h_p \, n_q \quad .
\end{eqnarray}

In Eqs.~(\ref{13})-(\ref{23}), $\nu_p$ equals one when the $p$th NO belongs to the set of the $N$ natural spinorbitals with the greatest occupation numbers (where $N$ is the number of electrons); otherwise it equals zero.  The other quantities are defined as
\begin{eqnarray}\label{24}
\eta_p = 1- \nu_p \quad , \quad h_p =1 - n_p \quad , \quad s_F =\sum_p \, \eta_p \, n_p  \quad .
\end{eqnarray}

Two comments are in order here.  First, because of its pairwise coupling of occupation numbers, the PNOF5 approach \cite{g7} is not included in the present study.  Second, the presence of single-index terms in the expressions for $U$ pertinent to spin-unpolarized systems implies the lack of invariance with respect to unitary transformations among NOs with degenerate occupation numbers -- the same problem that afflicts some of the functionals of the first category.

\vspace{16pt}

\normalsize\noindent\underline{III. DETAILS OF CALCULATIONS}\vspace{8pt}

The set of validation tools employed in the present work comprises the lowest-energy $^2P_-$ and $^4P_+$ states of the three-electron harmonium atom and the lowest-energy  $^1D_+$,  $^3P_+$, and 
$^5S_-$ states of the four-electron species in combination with 19 magnitudes of the confinement strength $\omega$ that span the range from $10^{-3}$ to $10^3$ \cite{y1,y2}.  The three-electron wavefunctions of the $^2P_-$ and $^4P_+$ symmetries describe, respectively, the ground and the first excited states for all confinement strengths.  On the other hand, the ground state of the four-electron harmonium atom has the  $^3P_+$ symmetry for large values of $\omega$ and $^5S_-$ for small ones, the transition occurring at $\omega\,\approx 0.0240919$ \cite{y2}.  The $^1D_+$ state always lies above its $^3P_+$ counterpart.
\vspace{6pt}

\begin{flushright} TABLES I-III HERE \end{flushright}

\vspace{6pt}
In the case of the $^2P_-$ and $^4P_+$ states of the three-electron harmonium atom, the values of the correlation component $U(\omega)$ of the electron-electron repulsion energy accurate to within a few $\mu$hartree have been published elsewhere \cite{z}.  Application of the same numerical methods to the results of high-quality electronic structure calculations on the four-electron species \cite{y2} produces the data listed in Tables I-III.  The computation of the NOs and their occupation numbers has been described previously \cite{y1,y2}. 

At the $\omega \to \infty$ limit of vanishing electron correlation, the values of $U(\omega)$ tend to constants that equal \cite{t,z}
\begin{eqnarray}\label{25}
-\frac{98}{135}+\frac{967}{135 \, \pi}-\frac{14 \, \sqrt{3}}{45 \, \pi}+\frac{43 \, \ln 2}{\pi}
-\frac{310 \, \ln (1+\sqrt{3})}{9 \, \pi} \approx -0.149 \, 481 \, 001 \; ,
\end{eqnarray}
\begin{eqnarray}\label{26}
\frac{2}{15}+\frac{64}{15 \, \pi}-\frac{16 \, \sqrt{3}}{15 \, \pi}+\frac{80 \, \ln 2}{3 \, \pi}
-\frac{64 \, \ln (1+\sqrt{3})}{3 \, \pi} \approx -0.037 \, 933 \, 367 \; ,
\end{eqnarray}
\begin{eqnarray}\label{27}
-\frac{4}{5}+\frac{1612}{135 \, \pi}-\frac{44 \, \sqrt{3}}{45 \, \pi}+\frac{674 \, \ln 2}{9 \, \pi}
-\frac{60 \, \ln (1+\sqrt{3})}{\pi} \approx -0.210 \, 155 \, 924 \; ,
\end{eqnarray}
and
\begin{eqnarray}\label{28}
\frac{4}{9}+\frac{80}{9 \, \pi}-\frac{8 \, \sqrt{3}}{3 \, \pi}+\frac{160 \, \ln 2}{3 \, \pi}
-\frac{128 \, \ln (1+\sqrt{3})}{3 \, \pi} \approx -0.078 \, 954 \, 185 \; ,
\end{eqnarray}
for the $^2P_-$, $^4P_+$, $^3P_+$, and $^5S_-$ states, respectively, reflecting the singly-determinantal nature of the underlying wavefunctions.  The fidelity with which these constants are reproduced reflects the performance of approximate 1-matrix functionals for systems of diverse spin multiplicities in which only the dynamical electron correlation is present.

In contrast, faithful reproduction of the analogous asymptotics for the multideterminantal $^1D_+$ state, which reads $U(\omega) = -\frac{1}{3} \, \sqrt{\frac{2}{\pi}} \, \sqrt{\omega} + \dots \;$, indicates the suitability of a given functional for description of systems with the nondynamical electron correlation due to degeneracy of the zeroth-order wavefunction.  Finally, the differences between the exact and computed values of $U(\omega)$ within the strong-correlation regime of small $\omega$ quantify the performance of approximate functionals for quasi-classical systems that lack strongly occupied NOs \cite{u,x}.

For the approximate expressions (\ref{4})-(\ref{12}), the validation encompasses all the five states in question.  On the other hand, for the functionals of the PNOF family, the test runs involve only the  $^4P_+$,  $^1D_+$, and $^5S_-$ states due to the aforediscussed restrictions on the spin multiplicity.
\vspace{6pt}

\begin{flushright} TABLES IV AND V HERE \end{flushright}

\vspace{4pt}
\normalsize\noindent\underline{IV. RESULTS AND DISCUSSION}\vspace{8pt}

Several dichotomies are discernible in the performances of the 1-matrix functionals for the  $^2P_-$ ground state of the three-electron harmonium atom (Table IV).  Within the weak-correlation regime, the MBB, GU, BBC1, and BBC2 functionals  reproduce the exact values of $U$ with comparable accuracy.  What sets them apart, however, is the ability to correctly describe the strongly correlated species.  Here, the GU functional fares the best, whereas the BBC1 and BBC2 estimates for $U$ exhibit a sudden drop in accuracy for $\omega < 10^{-2}$ due to the "phase-switching" step functions present in Eqs.~(\ref{8}) and (\ref{9}).  The MBB, CA, and CGA expressions yield similarly poor approximations to $U$.  Paradoxically, the performance  of the ML and ML-SIC functionals, which is very unsatisfactory at the weak-correlation limit, improves significantly upon weakening of the confinement, making them the most accurate ones at the smallest values of $\omega$. 

Inspection of the data compiled in Table V reveals an entirely different situation for the $^4P_+$ high-spin state where none of the functionals involving solely the exchange integrals is capable of producing reasonable approximations to $U$ as $\omega \to \infty$.  Their performance for the strongly correlated species is somewhat better with only the BBC1, BBC2, ML, and ML-SIC expressions being reasonably accurate. 
\vspace{6pt}

\begin{flushright} TABLE VI HERE \end{flushright}

\vspace{6pt}
Due to its multi-determinantal character, the  $^1D_+$ state of the four-electron harmonium atom turns out to be particularly challenging for the "$K$-only" functionals.  In fact, a rough agreement with the exact data is observed only for the GU, ML, and ML-SIC approximate values of $U$ within the strong-correlation regime (Table VI).  The positive-valuedness of the BBC1 and BBC2 estimates for small $\omega$ is worth noting in this context. 
\vspace{6pt}

\begin{flushright} TABLES VII AND VIII HERE \end{flushright}

\vspace{6pt}
According to the data displayed in Tables VII and VIII, the performance of the functionals under study for the $^3P_+$ and $^5S_-$ states of the four-electron species largely parallels that for the $^2P_-$ and $^4P_+$ states of its three-electron counterpart.  Thus, within the weak-correlation regime of the $^3P_+$ state, the MBB, GU, BBC1, and BBC2 functionals are again the most accurate (though somewhat less than in the analogous case of the $^2P_-$ state). Similarly, the ML and ML-SIC functionals perform well only for very small values of $\omega$.  At that limit, they are more accurate then the GU approximation, whereas the  MBB, CA, and CGA expressions again yield similarly poor estimates of $U$. The BBC1 and BBC2 functionals fare the worst, producing positive values of $U$.

Not surprisingly, the results of test calculations for the $^5S_-$ high-spin state reveal extremely poor reproduction of the exact correlation components of the electron-electron repulsion energies by all the functionals, the only exception being the ML and ML-SIC ones at small confinement strengths.  In contrast to the case of the $^4P_+$ state, the BBC1 and BBC2 expressions yield positive values of $U$ at the $\omega \to 0$ limit. 
\vspace{6pt}

\begin{flushright} TABLES IX - XI HERE \end{flushright}

\vspace{6pt}
The data compiled in Tables IX-XI uncover a substantial underestimation or even a complete neglect of the electron correlation effects in high-spin states by the functionals of the PNOF family.  For the PNOF1, PNOF2, and PNOF3 ones, this flaw persists for the $^1D_+$ state.  On the other hand, among those included in the present survey, the PNOF4 and PNOF6 functionals are the only ones capable of describing the $\omega \to \infty$ limit of this state with decent accuracy.  Although their performance quickly deteriorates upon weakening of the confinement, the latter functional yields estimates of $U$ that are both negative and greater than the exact ones for all values of $\omega$ (Table X).

Overall, one finds the functionals under study disappointingly inaccurate.  With the exception of the PNOF4 and PNOF6 ones, all of them fail for a system described by a wavefunction with two predominant Slater determinants.  Moreover, none of the functionals performs satisfactorily for high-spin species within the weak-correlation regime, signaling an unbalanced description of systems with different spin multiplicities.  At the strong-correlation limit, only the ML and ML-SIC expressions appear to perform reasonably well.  However, it is not clear at this point whether their accuracy is not purely accidental in this instance.

Most of the flaws and failures uncovered by the present survey have readily traceable origins.  The most obvious one is the presence of the "phase-switching" step functions in the BBC1 and BBC2 expressions.  Whereas seemingly improving description of dissociation limits of simple molecules with single bonds \cite{f7}, it introduces discontinuities in the first-order derivatives of $U(\omega)$ with respect to $\omega$ and results in the quick deterioration of accuracy apparent upon further lowering of the confinement strength.

The poor performance of the ML and ML-SIC functionals for weakly correlated systems can be elucidated with equal ease.  In such systems, $U(\omega)$ tends to a constant at the limit of $\omega \to \infty$ [see Eqs.~(\ref{25})-(\ref{28})], whereas the electron-electron repulsion integrals grow like $\omega^{1/2}$.  Consequently, the expressions premultiplying these integrals have to be asymptotically proportional to $\omega^{-1/2}$.  The deviations of the occupation numbers $\{ n_p \}$ from their $\omega \to \infty$ limiting values (equal to either 0 or 1) scale asymptotically like $\omega^{-1}$ \cite{l,jc1}.  Hence, the leading asymptotic terms of the premultipliers have to be proportional to square roots of the products $\{ n_p \, n_q \}$ for the pairs comprising one weakly and one strongly occupied NO that exclusively contribute to $U(\omega)$ at this limit \cite{jc2}.  Obviously, the expressions (\ref{10}) and (\ref{11}) do not conform to this constraint.
\vspace{-5pt}

\begin{flushright} TABLE XII HERE \end{flushright}

\vspace{6pt}
This conclusion is confirmed by the trends in the exponents $\lambda$ of the power functional (\ref{12}) individually optimized for each state and $\omega$ (Table XII).  The convergence to the limiting value of $\frac{1}{2}$ is clearly discernible for all the singly-determinantal cases, its rate strongly depending on the state in question.  An analogous analysis for the $^1D_+$ state yields the leading $\omega \to \infty$ asymptotics of $\frac{10 \sqrt{2}+3}{30}$, $\frac{40 \sqrt{2}-37}{120}$, $\frac{13}{30}$, $\frac{5 \sqrt{3}+8}{30}$, $\frac{\sqrt{2}}{3}$, and $\frac{\sqrt{2}}{3}$ (all in the units of 
$-\sqrt{\frac{2}{\pi}} \, \sqrt{\omega}$) for $U$ afforded by the MBB, GU, CA, GCA, BBC1, and BBC2 functionals, respectively.  These asymptotics, which are compatible with the large-$\omega$ entries in Table VI, should be compared with the exact limit of $\frac{1}{3}$ mentioned in the previous section of this paper.  The limiting value of $\lambda$ turns out to be given by the rather unwieldy expression
$\frac{\ln (2 \, \sqrt{454}+10)-\ln 33}{\ln 2} \approx 0.672 \, 996$ that compares well with the actual data computed within the weak-correlation regime (Table XII).

Finally, a comment on the uneven performance of the functionals for systems with diverse spin multiplicities is in order.   This problem stems from the insufficiency of the occupation numbers alone for a proper reconstruction of the 2-cumulant matrix that is equally applicable to species with balanced and unbalanced spins \cite{jc3}.

\vspace{16pt}

\normalsize\noindent\underline{V. CONCLUSIONS}\vspace{8pt}

A simple comparison between the exact and approximate correlation components $U$ of the electron-electron repulsion energy of several states of few-electron harmonium atoms with varying confinement strengths provides a stringent validation tool for 1-matrix functionals. The robustness of this tool is clearly demonstrated in a survey of 14 known functionals, which reveals their substandard performance within different electron correlation regimes.  Unlike spot-testing that employs dissociation curves of diatomic molecules or more extensive benchmarking against experimental atomization energies of molecules comprising some standard set, the present approach not only uncovers the flaws and patent failures of the functionals but, even more importantly, also allows for pinpointing their root causes.  Since the approximate values of $U$ are computed at exact 1-densities, the testing requires minimal programming,  
and thus is particularly suitable for rapid screening of new functionals.

The conclusion that emerges from the survey of functionals reported in this paper is that the current approximate incarnations of DMFT are highly unlikely to compete with the more traditional approaches to the electron correlation problem.  However, the present study clearly identifies the deficiencies that have to be rectified and provides obvious clues for the direction of the future work.

\vspace{16pt}
\normalsize\noindent\underline{ACKNOWLEDGMENT}
\vspace{6pt}

The research described in this publication has been funded by NCN (Poland) under grant DEC-2012/07/B/ST4/00553, the MINECO projects CTQ2012-38496-C05-01 and CTQ2014-52525-P, and the Basque Country Consolidated Group Project No. IT588-13.  The authors gratefully acknowledge the computational resources granted at the MareNostrum computer of the Barcelona Supercomputing Center,
and the technical and human support provided by SGI/IZO-SGIker UPV/EHU.

\newpage

\noindent
Table I. The Coulomb, exchange, and correlation components of the electron-electron repulsion energies of the $^1D_+$ state of the four-electron harmonium atom.$^{a)}$

\vspace{12pt}
\noindent
\begin{tabular}{cccc}
\hline
\noalign{\vspace{2pt}}
$\omega$ &	$J_{\alpha\alpha}(\omega)=J_{\beta\beta}(\omega) =J_{\alpha\beta}(\omega)=J_{\beta\alpha}(\omega)$ &	$K_{\alpha\alpha}(\omega)=K_{\beta\beta}(\omega)$ &	$U(\omega)$\\
\hline
1000 & 43.141266 & -22.197854 & -8.636071	\\
500 & 30.455349 & -15.669914 & -6.172181	\\
200 & 19.199171 & -9.877439 & -3.985323	\\
100 & 13.526552 & -6.958008 & -2.882509	\\
50 & 9.516011 & -4.893624 & -2.101896	\\
20 & 5.958643 & -3.061782 & -1.407503	\\
10 & 4.167172 & -2.138516 & -1.055637	\\
5 & 2.902184 & -1.485702 & -0.804504	\\
2 & 1.783379 & -0.906700 & -0.576873	\\
1 & 1.223264 & -0.615423 & -0.457180	\\
0.5 & 0.831403 & -0.410474 & -0.366884	\\
0.2 & 0.491092 & -0.231455 & -0.276380	\\
0.1 & 0.325705 & -0.144669 & -0.221710	\\
0.05 & 0.213954 & -0.087269 & -0.174972	\\
0.02 & 0.121373 & -0.042626 & -0.122989	\\
0.01 & 0.078588 & -0.024254 & -0.090860	\\
0.005 & 0.050715 & -0.013733 & -0.065122	\\
0.002 & 0.028306 & -0.006503 & -0.040335	\\
0.001 & 0.018158 & -0.003713 & -0.027465	\\

\hline
\end{tabular}
\\\\
\noindent$^{a)}$ All values in hartree.

\newpage
\noindent
Table II. The Coulomb, exchange, and correlation components of the electron-electron repulsion energies of the $^3P_+$ state of the four-electron harmonium atom.$^{a)}$

\vspace{12pt}
\noindent
\begin{tabular}{ccccccc}
\hline
\noalign{\vspace{2pt}}
$\omega$ &	$J_{\alpha\alpha}(\omega)$ & $J_{\beta\beta}(\omega) $ & $J_{\alpha\beta}(\omega)$ &	$K_{\alpha\alpha}(\omega)$ & $K_{\beta\beta}(\omega)$ &	$U(\omega)$\\
\hline
1000 & 93.008124 & 12.560742 & 67.008664 & -42.746677 & -12.560229 & -0.209290	\\
500 & 65.665670 & 8.865790 & 47.302350 & -30.183562 & -8.865067 & -0.208932	\\
200 & 41.404798 & 5.587243 & 29.816961 & -19.035959 & -5.586108 & -0.208226	\\
100 & 29.178390 & 3.934970 & 21.005052 & -13.417624 & -3.933378 & -0.207436	\\
50 & 20.534329 & 2.766778 & 14.774985 & -9.444926 & -2.764550 & -0.206329	\\
20 & 12.867011 & 1.730514 & 9.248786 & -5.919992 & -1.727066 & -0.204168	\\
10 & 9.005784 & 1.208582 & 6.465715 & -4.143653 & -1.203827 & -0.201780	\\
5 & 6.279272 & 0.839956 & 4.500418 & -2.887936 & -0.833470 & -0.198484	\\
2 & 3.867700 & 0.513792 & 2.762001 & -1.774624 & -0.504265 & -0.192192	\\
1 & 2.660140 & 0.350409 & 1.891486 & -1.214758 & -0.338043 & -0.185436	\\
0.5 & 1.814915 & 0.236077 & 1.282302 & -0.820685 & -0.220637 & -0.176396	\\
0.2 & 1.079828 & 0.136943 & 0.753144 & -0.475249 & -0.117985 & -0.159978	\\
0.1 & 0.721360 & 0.089136 & 0.496129 & -0.305781 & -0.068972 & -0.143613	\\
0.05 & 0.477742 & 0.057401 & 0.322946 & -0.191126 & -0.038073 & -0.123918	\\
0.02 & 0.273685 & 0.032044 & 0.180723 & -0.098028 & -0.016717 & -0.094706	\\
0.01 & 0.178013 & 0.020771 & 0.116010 & -0.057445 & -0.009267 & -0.073032	\\
0.005 & 0.115072 & 0.013511 & 0.074404 & -0.033092 & -0.005332 & -0.054130	\\
0.002 & 0.064224 & 0.007628 & 0.041282 & -0.015704 & -0.002604 & -0.034834	\\
0.001 & 0.041182 & 0.004928 & 0.026380 & -0.008877 & -0.001504 & -0.024313	\\

\hline
\end{tabular}
\\\\
\noindent$^{a)}$ All values in hartree. The entries labeled $J_{\alpha\beta}(\omega)$ actually list the sums of two equal contributions $J_{\alpha\beta}(\omega)$ and $J_{\beta\alpha}(\omega)$.

\newpage
\noindent
Table III. The Coulomb, exchange, and correlation components of the electron-electron repulsion energies of the $^5S_-$ state of the four-electron harmonium atom.$^{a)}$

\vspace{12pt}
\begin{tabular}{cccc}
\hline
\noalign{\vspace{2pt}}
$\omega$ & $J_{\alpha\alpha}(\omega)$ & $K_{\alpha\alpha}(\omega)$ & $U(\omega)$ \\
\hline
1000 & 160.375791 & -59.760718 & -0.078815	\\
500 & 113.264846 & -42.209350 & -0.078755	\\
200 & 71.462766 & -26.635569 & -0.078639	\\
100 & 50.395597 & -18.786507 & -0.078511	\\
50 & 35.500157 & -13.236518 & -0.078326	\\
20 & 22.285777 & -8.312150 & -0.077967	\\
10 & 15.628837 & -5.830637 & -0.077563	\\
5 & 10.925499 & -4.076432 & -0.076996	\\
2 & 6.760058 & -2.521016 & -0.075881	\\
1 & 4.669020 & -1.738458 & -0.074639	\\
0.5 & 3.200003 & -1.186941 & -0.072903	\\
0.2 & 1.914169 & -0.701637 & -0.069530	\\
0.1 & 1.281760 & -0.461413 & -0.065868	\\
0.05 & 0.849028 & -0.296450 & -0.061013	\\
0.02 & 0.485195 & -0.158408 & -0.052641	\\
0.01 & 0.314876 & -0.095325 & -0.045071	\\
0.005 & 0.203205 & -0.055810 & -0.036971	\\
0.002 & 0.113236 & -0.026611 & -0.026582	\\
0.001 & 0.072535 & -0.014975 & -0.019736	\\
\hline
\end{tabular}
\\\\
\noindent$^{a)}$ All values in hartree.

\newpage
\noindent
Table IV. The  correlation components of the electron-electron repulsion \\ energies of the $^2P_-$ state of the three-electron harmonium atom.

\vspace{12pt}
\begin{tabular}{rrrrrrrrrr}
\hline
\noalign{\vspace{2pt}}
\multicolumn{1}{c}{\multirow{2}{*}{$\omega$}} & \multicolumn{9}{c}{$U(\omega)$ [mhartree]} \\
\cmidrule{2-10}
& exact & MBB & GU & CA & CGA & BBC1 & BBC2 & ML & ML-SIC \\
\hline
1000 & -149.0 & -151.8 & -150.6 & -2.3 & -107.9 & -150.9 & -150.9 & -33.3 & -293.5	\\
500 & -148.8 & -152.1 & -150.4 & -3.2 & -108.3 & -150.9 & -150.8 & -46.3 & -360.2	\\
200 & -148.4 & -152.6 & -150.0 & -5.1 & -109.1 & -150.7 & -150.5 & -70.4 & -415.5	\\
100 & -148.0 & -153.1 & -149.5 & -7.1 & -110.0 & -150.4 & -150.2 & -93.6 & -415.6	\\
50 & -147.4 & -153.8 & -148.6 & -9.9 & -111.1 & -150.0 & -149.7 & -119.0 & -383.2	\\
20 & -146.2 & -154.9 & -146.9 & -15.4 & -113.2 & -149.0 & -148.7 & -146.5 & -315.5	\\
10 & -144.9 & -155.9 & -145.0 & -21.2 & -115.4 & -147.9 & -147.4 & -154.2 & -261.9	\\
5 & -143.0 & -157.0 & -142.1 & -28.8 & -118.0 & -146.2 & -145.5 & -148.3 & -213.6	\\
2 & -139.3	& -158.2 & -136.3 & -42.3 & -122.3 & -142.7 & -141.6 & -127.2 & -162.1	\\
1 & -135.1 & -158.3 & -129.7 & -54.8 & -125.6 & -138.6 & -137.1 & -110.4 & -133.3	\\
0.5 & -129.2 & -156.7 & -120.8 & -68.4 & -128.1 & -132.5 & -130.6 & -98.9 & -110.8	\\
0.2 & -117.9 & -149.6 & -104.7 & -84.8 & -127.7 & -120.4 & -117.9 & -89.6 & -85.5	\\
0.1 & -106.1 & -138.9 & -90.0 & -92.1 & -122.6 & -107.4 & -104.5 & -79.5 & -68.1	\\
0.05 & -91.6 & -123.4 & -74.4 & -92.6 & -112.3 & -91.1 & -88.0 & -66.6 & -51.9	\\
0.02 & -69.9 & -99.4 & -56.7 & -83.6 & -93.4 & -66.1 & -63.1 & -50.6 & -37.5	\\
0.01 & -53.9 & -81.3 & -46.3 & -72.2 & -77.7 & -47.3 & -44.6 & -41.0 & -32.2	\\
0.005 & -40.0 & -64.2 & -37.3 & -59.1 & -62.1 & -14.9 & -12.6 & -33.2 & -27.6	\\
0.002 & -25.7 & -45.2 & -27.4 & -42.9 & -44.2 & -6.3 & -4.7 & -24.9 & -22.1	\\
0.001 & -17.9 & -33.9 & -21.4 & -32.6 & -33.3 & -2.1 & -0.9 & -19.8 & -18.4	\\
\hline
\end{tabular}

\newpage
\noindent
Table V. The  correlation components of the electron-electron repulsion \\ energies of the $^4P_+$ state of the three-electron harmonium atom.

\vspace{12pt}
\begin{tabular}{rrrrrrrrrr}
\hline
\noalign{\vspace{2pt}}
\multicolumn{1}{c}{\multirow{2}{*}{$\omega$}} & \multicolumn{9}{c}{$U(\omega)$ [mhartree]} \\
\cmidrule{2-10}
& exact & MBB & GU & CA & CGA & BBC1 & BBC2 & ML & ML-SIC \\
\hline
1000 & -37.9 & -82.8 & -82.6 & -0.6 & -58.7 & -82.6 & -82.6 & -12.3 & -117.6	\\
500 & -37.9 & -82.9 & -82.5 & -0.8 & -58.8 & -82.6 & -82.6 & -17.2 & -154.2	\\
200 & -37.9 & -82.9 & -82.3 & -1.3 & -59.0 & -82.5 & -82.4 & -26.7 & -202.0	\\
100 & -37.9 & -82.9 & -82.0 & -1.8 & -59.1 & -82.3 & -82.2 & -36.6 & -225.6	\\
50 & -37.9 & -82.9 & -81.7 & -2.6 & -59.2 & -82.1 & -81.9 & -48.9 & -230.2	\\
20 & -37.9 & -82.8 & -80.9 & -4.0 & -59.6 & -81.6 & -81.3 & -67.0 & -208.9	\\
10 & -37.9 & -82.7 & -80.1 & -5.5 & -59.9 & -81.0 & -80.6 & -78.5 & -180.9	\\
5 & -37.9 & -82.6 & -78.9 & -7.6 & -60.3 & -80.1 & -79.6 & -84.2 & -150.2	\\
2 & -37.9 & -82.1 & -76.6 & -11.5 & -60.9 & -78.4 & -77.6 & -80.6 & -113.2	\\
1 & -37.7 & -81.5 & -74.1 & -15.4 & -61.4 & -76.6 & -75.5 & -71.8 & -90.3	\\
0.5 & -37.5 & -80.4 & -70.7 & -20.1 & -61.9 & -74.1 & -72.6 & -61.6 & -72.4	\\
0.2 & -36.9 & -77.8 & -64.7 & -27.3 & -62.0 & -69.3 & -67.3 & -50.5 & -54.9	\\
0.1 & -35.9 & -74.6 & -58.9 & -32.8 & -61.3 & -64.4 & -61.8 & -44.7 & -44.5	\\
0.05 & -34.4 & -69.9 & -52.1 & -37.4 & -59.3 & -58.1 & -55.1 & -39.2 & -35.1	\\
0.02 & -31.0 & -61.1 & -42.1 & -40.4 & -54.0 & -47.8 & -44.5 & -31.2 & -23.9	\\
0.01 & -27.4 & -53.9 & -35.7 & -40.2 & -49.1 & -39.1 & -35.9 & -25.7 & -19.7	\\
0.005 & -23.1 & -46.3 & -30.0 & -37.7 & -43.1 & -30.1 & -27.1 & -21.4 & -17.4	\\
0.002 & -17.2 & -35.7 & -23.2 & -31.5 & -34.1 & -18.9 & -16.5 & -16.6 & -14.3	\\
0.001 & -13.0 & -28.3 & -18.7 & -25.9 & -27.3 & -12.0 & -10.2 & -13.5 & -12.1	\\
\hline
\end{tabular}

\newpage
\noindent
Table VI. The  correlation components of the electron-electron repulsion \\ energies of the $^1D_+$ state of the four-electron harmonium atom.

\vspace{12pt}
\hspace*{-7pt}\makebox[\linewidth][c]{
\begin{tabular}{rrrrrrrrrr}
\hline
\noalign{\vspace{2pt}}
\multicolumn{1}{c}{\multirow{2}{*}{$\omega$}} & \multicolumn{9}{c}{$U(\omega)$ [mhartree]} \\
\cmidrule{2-10}
& exact & MBB & GU & CA & CGA & BBC1 & BBC2 & ML & ML-SIC \\
\hline
1000 & -8636.0 & -14600.8 & -4334.2 & -11024.8 & -14149.9 & -11824.6 & -11824.6 & -2239.1 & -738.6	\\
500 & -6172.2 & -10378.4 & -3128.8 & -7823.2 & -10046.3 & -8341.2 & -8341.2 & -1609.7 & -725.3	\\
200 & -3985.3 & -6631.9 & -2058.8 & -4983.2 & -6405.6 & -5250.7 & -5250.7 & -1066.8 & -730.9	\\
100 & -2882.6 & -4743.9 & -1519.0 & -3553.3 & -4571.1 & -3693.4 & -3693.4 & -807.8 & -717.1	\\
50 & -2101.9 & -3409.1 & -1136.7 & -2542.7 & -3274.4 & -2592.8 & -2592.8 & -637.4 & -672.5	\\
20 & -1407.5 & -2225.0 & -795.9 & -1648.2 & -2124.9 & -1617.1 & -1617.1 & -497.9 & -572.9	\\
10 & -1055.6 & -1628.5 & -622.4 & -1199.6 & -1546.5 & -1126.5 & -1126.5 & -426.4 & -484.4	\\
5 & -804.5 & -1206.7 & -497.5 & -885.0 & -1138.5 & -780.9 & -780.9 & -365.9 & -399.1	\\
2 & -576.9 & -831.9 & -381.8 & -610.1 & -777.8 & -476.8 & -476.8 & -291.8 & -303.7	\\
1 & -457.2 & -641.6 & -318.2 & -474.8 & -596.6 & -326.0 & -326.0 & -242.2 & -247.7	\\
0.5 & -366.9 & -504.3 & -267.0 & -381.1 & -467.7 & -221.8 & -221.8 & -202.6 & -202.6	\\
0.2 & -276.4 & -375.1 & -210.6 & -297.3 & -349.3 & -132.6 & -132.6 & -162.4 & -153.1	\\
0.1 & -221.7 & -302.0 & -173.3 & -250.7 & -283.7 & -88.8 & -88.8 & -135.8 & -122.3	\\
0.05 & -175.0 & -242.1 & -140.4 & -210.8 & -230.2 & -57.5 & -57.5 & -111.6 & -96.7	\\
0.02 & -123.0 & -177.7 & -104.9 & -163.1 & -171.7 & -26.7 & -26.7 & -84.6 & -71.7	\\
0.01 & -90.9 & -138.0 & -83.8 & -130.2 & -134.6 & -9.5 & -9.5 & -68.5 & -59.2	\\
0.005 & -65.1 & -104.8 & -65.9 & -100.7 & -103.0 & 53.1 & 53.1 & -55.2 & -49.1	\\
0.002 & -40.3 & -70.7 & -46.8 & -68.9 & -69.9 & 36.9 & 36.9 & -40.8 & -37.8	\\
0.001 & -27.5 & -51.7 & -35.5 & -50.7 & -51.2 & 28.0 & 28.0 & -32.2 & -31.0	\\
\hline
\end{tabular}
}

\newpage
\noindent
Table VII. The  correlation components of the electron-electron repulsion \\ energies of the $^3P_+$ state of the four-electron harmonium atom.

\vspace{12pt}
\begin{tabular}{rrrrrrrrrr}
\hline
\noalign{\vspace{2pt}}
\multicolumn{1}{c}{\multirow{2}{*}{$\omega$}} & \multicolumn{9}{c}{$U(\omega)$ [mhartree]} \\
\cmidrule{2-10}
& exact & MBB & GU & CA & CGA & BBC1 & BBC2 & ML & ML-SIC \\
\hline
1000 & -209.3 & -218.9 & -217.3 & -3.2 & -155.5 & -217.6 & -217.5 & -53.0 & -451.9	\\
500 & -209.0 & -219.2 & -217.0 & -4.5 & -156.1 & -217.4 & -217.2 & -73.6 & -541.2	\\
200 & -208.2 & -219.7 & -216.3 & -7.1 & -157.1 & -216.9 & -216.6 & -110.7 & -600.6	\\
100 & -207.4 & -220.2 & -215.4 & -10.0 & -158.1 & -216.3 & -215.9 & -145.5 & -586.0	\\
50 & -206.3 & -220.8 & -214.1 & -13.9 & -159.5 & -215.3 & -214.7 & -181.2 & -532.4	\\
20 & -204.2 & -221.8 & -211.4 & -21.5 & -162.0 & -213.3 & -212.4 & -214.9 & -435.8	\\
10 & -201.8 & -222.6 & -208.3 & -29.6 & -164.6 & -210.9 & -209.7 & -219.9 & -361.9	\\
5 & -198.5 & -223.5 & -204.0 & -40.3 & -167.8 & -207.5 & -205.9 & -207.1 & -295.7	\\
2 & -192.2 & -224.0 & -195.5 & -59.0 & -173.0 & -200.8 & -198.3 & -176.3 & -225.8	\\
1 & -185.4 & -223.3 & -186.3 & -76.4 & -176.9 & -193.1 & -190.0 & -154.1 & -187.7	\\
0.5 & -176.4 & -220.3 & -174.2 & -95.5 & -180.0 & -182.5 & -178.5 & -139.1 & -159.0	\\
0.2 & -160.0 & -210.3 & -153.0 & -119.0 & -179.6 & -162.6 & -157.4 & -124.4 & -127.2	\\
0.1 & -143.6 & -196.0 & -133.6 & -130.5 & -173.2 & -142.4 & -136.5 & -110.0 & -103.7	\\
0.05 & -123.9 & -175.4 & -113.0 & -132.7 & -159.9 & -118.2 & -111.9 & -93.2 & -81.7	\\
0.02 & -94.7 & -141.6 & -87.4 & -120.4 & -133.5 & -82.4 & -76.5 & -71.2 & -59.8	\\
0.01 & -73.0 & -115.0 & -70.9 & -103.3 & -110.3 & -33.0 & -27.9 & -57.2 & -48.7	\\
0.005 & -54.1 & -90.4 & -56.8 & -84.0 & -87.8 & -19.7 & -15.8 & -45.8 & -40.3	\\
0.002 & -34.8 & -63.4 & -41.4 & -60.6 & -62.2 & 16.4 & 16.4 & -34.2 & -31.3	\\
0.001 & -24.3 & -47.5 & -32.1 & -46.0 & -46.8 & 28.0 & 28.0 & -27.3 & -26.0	\\
\hline
\end{tabular}

\newpage
\noindent
Table VIII. The  correlation components of the electron-electron repulsion \\ energies of the $^5S_-$ state of the four-electron harmonium atom.

\vspace{12pt}
\begin{tabular}{rrrrrrrrrr}
\hline
\noalign{\vspace{2pt}}
\multicolumn{1}{c}{\multirow{2}{*}{$\omega$}} & \multicolumn{9}{c}{$U(\omega)$ [mhartree]} \\
\cmidrule{2-10}
& exact & MBB & GU & CA & CGA & BBC1 & BBC2 & ML & ML-SIC \\
\hline
1000 & -78.8 & -150.9 & -150.3 & -1.3 & -107.0 & -150.5 & -150.4 & -30.5 & -276.1	\\
500 & -78.8 & -150.9 & -150.0 & -1.8 & -107.1 & -150.3 & -150.1 & -42.6 & -344.3	\\
200 & -78.6 & -150.8 & -149.5 & -2.8 & -107.3 & -149.9 & -149.7 & -65.1 & -407.2	\\
100 & -78.5 & -150.7 & -148.9 & -3.9 & -107.5 & -149.4 & -149.1 & -87.4 & -414.9	\\
50 & -78.3 & -150.5 & -148.0 & -5.4 & -107.8 & -148.7 & -148.3 & -112.5 & -390.0	\\
20 & -78.0 & -150.1 & -146.3 & -8.4 & -108.2 & -147.4 & -146.7 & -141.9 & -330.9	\\
10 & -77.6 & -149.6 & -144.3 & -11.6 & -108.7 & -145.8 & -144.9 & -152.4 & -280.4	\\
5 & -77.0 & -148.8 & -141.5 & -15.9 & -109.2 & -143.7 & -142.4 & -149.2 & -231.1	\\
2 & -75.9 & -147.2 & -136.3 & -23.6 & -110.0 & -139.5 & -137.5 & -130.8 & -173.8	\\
1 & -74.6 & -145.1 & -130.8 & -31.1 & -110.5 & -134.9 & -132.4 & -113.7 & -139.3	\\
0.5 & -72.9 & -141.9 & -123.6 & -40.0 & -110.6 & -128.7 & -125.5 & -98.7 & -112.4	\\
0.2 & -69.5 & -135.3 & -111.4 & -52.8 & -109.4 & -117.5 & -113.2 & -83.6 & -87.0	\\
0.1 & -65.9 & -127.7 & -100.1 & -61.8 & -106.6 & -106.4 & -101.3 & -72.5 & -71.5	\\
0.05 & -61.0 & -117.6 & -87.5 & -68.4 & -101.4 & -93.0 & -87.2 & -61.0 & -56.4	\\
0.02 & -52.6 & -100.5 & -70.3 & -71.0 & -90.4 & -72.6 & -66.6 & -47.9 & -40.0	\\
0.01 & -45.1 & -86.6 & -58.8 & -68.3 & -80.1 & -56.4 & -50.8 & -39.7 & -32.7	\\
0.005 & -37.0 & -72.6 & -48.8 & -62.0 & -68.6 & -40.5 & -35.7 & -32.7 & -27.9	\\
0.002 & -26.6 & -54.7 & -37.3 & -50.1 & -52.8 & -22.4 & -19.0 & -25.3 & -22.4	\\
0.001 & -19.7 & -42.7 & -29.8 & -40.3 & -41.6 & 16.3 & 16.3 & -20.8 & -19.0	\\
\hline
\end{tabular}

\newpage
\noindent
Table IX. The  correlation components of the electron-electron repulsion \\ energies of the $^4P_+$ state of the three-electron harmonium atom.

\vspace{12pt}
\begin{tabular}{rrrrrrr}
\hline
\noalign{\vspace{2pt}}
\multicolumn{1}{c}{\multirow{2}{*}{$\omega$}} & \multicolumn{6}{c}{$U(\omega)$ [mhartree]} \\
\cmidrule{2-7}
& exact & PNOF1 & PNOF2 & PNOF3 & PNOF4 & PNOF6  \\
\hline
1000 & -37.9 & -0.2 & -0.2 & 0.0 & -0.2 & -0.2	\\
500 & -37.9 & -0.3 & -0.3 & 0.0 & -0.3 & -0.3	\\
200 & -37.9 & -0.4 & -0.5 & 0.0 & -0.5 & -0.5	\\
100 & -37.9 & -0.7 & -0.7 & 0.0 & -0.7 & -0.7	\\
50 & -37.9 & -0.9 & -1.0 & 0.0 & -1.0 & -1.0	\\
20 & -37.9 & -1.4 & -1.5 & 0.0 & -1.5 & -1.5	\\
10 & -37.9 & -2.0 & -2.1 & 0.0 & -2.1 & -2.1	\\
5 & -37.9 & -2.8 & -2.9 & 0.0 & -2.9 & -2.9	\\
2 & -37.9 & -4.1 & -4.4 & 0.0 & -4.4 & -4.4	\\
1 & -37.7 & -5.5 & -5.9 & 0.0 & -5.9 & -5.9	\\
0.5 & -37.5 & -7.3 & -7.7 & 0.0 & -7.7 & -7.7	\\
0.2 & -36.9 & -9.8 & -10.5 & 0.0 & -10.5 & -10.5	\\
0.1 & -35.9 & -11.8 & -12.6 & 0.0 & -12.6 & -12.6	\\
0.05 & -34.4 & -13.4 & -14.3 & 0.0 & -14.3 & -14.2	\\
0.02 & -31.0 & -14.3 & -15.2 & 0.0 & -15.2 & -15.0	\\
0.01 & -27.4 & -13.7 & -14.7 & 0.0 & -14.7 & -14.3	\\
0.005 & -23.1 & -12.2 & -13.0 & 0.0 & -13.0 & -12.4	\\
0.002 & -17.2 & -9.4 & -9.9 & 0.0 & -9.9 & -9.0	\\
0.001 & -13.0 & -7.2 & -7.5 & 0.0 & -7.5 & -6.5	\\
\hline
\end{tabular}

\newpage
\noindent
Table X. The  correlation components of the electron-electron repulsion \\ energies of the $^1D_+$ state of the four-electron harmonium atom.

\vspace{12pt}
\begin{tabular}{rrrrrrr}
\hline
\noalign{\vspace{2pt}}
\multicolumn{1}{c}{\multirow{2}{*}{$\omega$}} & \multicolumn{6}{c}{$U(\omega)$ [mhartree]} \\
\cmidrule{2-7}
& exact & PNOF1 & PNOF2 & PNOF3 & PNOF4 & PNOF6  \\
\hline
1000 & -8636.0 & -4234.7 & -10707.7 & -4308.1 & -8446.2 & -8473.6	\\
500 & -6172.2 & -3029.1 & -7593.5 & -3102.4 & -5983.2 & -6010.5	\\
200 & -3985.3 & -1958.5 & -4830.2 & -2031.9 & -3798.0 & -3825.3	\\
100 & -2882.6 & -1418.8 & -3437.7 & -1492.2 & -2698.0 & -2725.3	\\
50 & -2101.9 & -1035.6 & -2452.8 & -1109.0 & -1919.7 & -1947.0	\\
20 & -1407.5 & -693.4 & -1578.6 & -766.7 & -1229.8 & -1257.1	\\
10 & -1055.6 & -518.6 & -1137.7 & -591.8 & -883.0 & -910.3	\\
5 & -804.5 & -392.3 & -825.6 & -465.2 & -638.8 & -665.9	\\
2 & -576.9 & -274.7 & -547.4 & -346.8 & -423.6 & -450.3	\\
1 & -457.2 & -210.2 & -405.7 & -281.2 & -316.2 & -342.1	\\
0.5 & -366.9 & -159.2 & -303.4 & -228.4 & -240.4 & -265.1	\\
0.2 & -276.4 & -105.8 & -207.8 & -171.1 & -171.3 & -193.0	\\
0.1 & -221.7 & -73.3 & -154.8 & -134.1 & -132.4 & -151.1	\\
0.05 & -175.0 & -46.5 & -112.4 & -101.7 & -99.3 & -114.6	\\
0.02 & -123.0 & -19.0 & -68.6 & -65.6 & -60.8 & -72.2	\\
0.01 & -90.9 & -4.0 & -43.8 & -43.5 & -36.2 & -45.9	\\
0.005 & -65.1 & 5.6 & -26.1 & -26.3 & -17.7 & -26.2	\\
0.002 & -40.3 & 11.4 & -12.0 & -11.1 & -2.8 & -10.3	\\
0.001 & -27.5 & 12.3 & -6.2 & -4.2 & 2.8 & -4.0	\\
\hline
\end{tabular}

\newpage
\noindent
Table XI. The  correlation components of the electron-electron repulsion \\ energies of the $^5S_-$ state of the four-electron harmonium atom.

\vspace{12pt}
\begin{tabular}{rrrrrrr}
\hline
\noalign{\vspace{2pt}}
\multicolumn{1}{c}{\multirow{2}{*}{$\omega$}} & \multicolumn{6}{c}{$U(\omega)$ [mhartree]} \\
\cmidrule{2-7}
& exact & PNOF1 & PNOF2 & PNOF3 & PNOF4 & PNOF6  \\
\hline
1000	&	-78.8	&	-0.5	&	-0.5	&	0.0	&	-0.5	&	-0.5	\\
500	&	-78.8	&	-0.6	&	-0.7	&	0.0	&	-0.7	&	-0.7	\\
200	&	-78.6	&	-0.9	&	-1.0	&	0.0	&	-1.0	&	-1.0	\\
100	&	-78.5	&	-1.3	&	-1.4	&	0.0	&	-1.4	&	-1.4	\\
50	&	-78.3	&	-1.9	&	-2.0	&	0.0	&	-2.0	&	-2.0	\\
20	&	-78.0	&	-2.9	&	-3.1	&	0.0	&	-3.1	&	-3.1	\\
10	&	-77.6	&	-4.0	&	-4.3	&	0.0	&	-4.3	&	-4.3	\\
5	&	-77.0	&	-5.5	&	-5.9	&	0.0	&	-5.9	&	-5.9	\\
2	&	-75.9	&	-8.1	&	-8.7	&	0.0	&	-8.7	&	-8.7	\\
1	&	-74.6	&	-10.7	&	-11.4	&	0.0	&	-11.4	&	-11.4	\\
0.5	&	-72.9	&	-13.7	&	-14.6	&	0.0	&	-14.6	&	-14.6	\\
0.2	&	-69.5	&	-17.9	&	-19.2	&	0.0	&	-19.2	&	-19.1	\\
0.1	&	-65.9	&	-20.7	&	-22.2	&	0.0	&	-22.2	&	-22.1	\\
0.05	&	-61.0	&	-22.5	&	-24.2	&	0.0	&	-24.2	&	-23.9	\\
0.02	&	-52.6	&	-22.6	&	-24.5	&	0.0	&	-24.5	&	-23.7	\\
0.01	&	-45.1	&	-20.9	&	-22.6	&	0.0	&	-22.6	&	-21.2	\\
0.005	&	-37.0	&	-17.8	&	-19.2	&	0.0	&	-19.2	&	-17.4	\\
0.002	&	-26.6	&	-13.1	&	-13.9	&	0.0	&	-13.9	&	-11.7	\\
0.001	&	-19.7	&	-9.7	&	-10.1	&	0.0	&	-10.1	&	-7.9	\\
\hline
\end{tabular}

\newpage
\noindent
Table XII. The  optimal exponents $\lambda$ in the power functional (\ref{12}).

\vspace{12pt}
\begin{tabular}{rccccc}
\hline
\noalign{\vspace{2pt}}
$\omega$ &$^2P_-$  & $^4P_+$   & $^1D_+$  & $^3P_+$   & $^5S_-$   \\
\hline
1000	&	0.5014	&	0.5561	&	0.6657	&	0.5034	&	0.5487	\\
500	&	0.5017	&	0.5590	&	0.6629	&	0.5039	&	0.5514	\\
200	&	0.5024	&	0.5634	&	0.6576	&	0.5046	&	0.5554	\\
100	&	0.5031	&	0.5671	&	0.6520	&	0.5055	&	0.5589	\\
50	&	0.5040	&	0.5712	&	0.6447	&	0.5066	&	0.5628	\\
20	&	0.5060	&	0.5774	&	0.6321	&	0.5088	&	0.5687	\\
10	&	0.5081	&	0.5827	&	0.6205	&	0.5111	&	0.5739	\\
5	&	0.5111	&	0.5886	&	0.6075	&	0.5143	&	0.5798	\\
2	&	0.5166	&	0.5973	&	0.5903	&	0.5203	&	0.5886	\\
1	&	0.5223	&	0.6044	&	0.5791	&	0.5265	&	0.5961	\\
0.5	&	0.5292	&	0.6119	&	0.5708	&	0.5342	&	0.6042	\\
0.2	&	0.5400	&	0.6218	&	0.5652	&	0.5464	&	0.6150	\\
0.1	&	0.5488	&	0.6287	&	0.5651	&	0.5566	&	0.6228	\\
0.05	&	0.5579	&	0.6345	&	0.5682	&	0.5671	&	0.6295	\\
0.02	&	0.5716	&	0.6402	&	0.5761	&	0.5815	&        0.6366	\\
0.01	&	0.5830	&	0.6446	&	0.5839	&	0.5923	&	0.6413	\\
0.005	&	0.5940	&	0.6492	&	0.5918	&	0.6024	&	0.6458	\\
0.002	&	0.6070	&	0.6551	&	0.6019	&	0.6142	&	0.6516	\\
0.001	&	0.6158	&	0.6595	&	0.6091	&	0.6222	&	0.6559	\\

\hline
\end{tabular}


\begin{thebibliography}{99}
\bibitem{cc} Due to the real-valuedness of the natural orbitals in question, no distinction is made in this paper between the exchange and the so-called L integrals.
\bibitem{f4} Strictly speaking, $U_{GU}$ is not a genuine 1-matrix functional as it is not invariant to unitary transformations among NOs with degenerate occupation numbers.
\bibitem{f8} Like $U_{GU}$, these expressions are not invariant to unitary transformations among NOs with degenerate occupation numbers.
\bibitem{jc2} This is a particular case of a general property of the "JKL-only" functionals \cite{jc3}.
\end{thebibliography}
\end{document}